Advancing 3D Medical Image Segmentation: Unleashing the Potential of Planarian Neural Networks in Artificial Intelligence

Ziyuan Huang[1, 2], Kevin Huggins[1], and Srikar Bellur[1]

[1]Harrisburg University of Science and Technology, Harrisburg, Pennsylvania, USA

[2]University of Massachusetts Chan Medical School, Worcester, Massachusetts, USA

**Author Note**

Correspondence concerning this article should be addressed to Ziyuan Huang, University of Massachusetts Chan Medical School, 368 Plantation Street, Worcester, MA 01605. Email: Ziyuan.Huang2@umassmed.edu

Advancing 3D Medical Image Segmentation: Unleashing the Potential of Planarian Neural Networks in Artificial Intelligence

**Abstract**

Our study presents PNN-UNet as a method for constructing deep neural networks that replicate the planarian neural network (PNN) structure in the context of 3D medical image data. Planarians typically have a cerebral structure comprising two neural cords, where the cerebrum acts as a coordinator, and the neural cords serve slightly different purposes within the organism's neurological system. Accordingly, PNN-UNet comprises a Deep-UNet and a Wide-UNet as the nerve cords, with a densely connected autoencoder performing the role of the brain. This distinct architecture offers advantages over both monolithic (UNet) and modular networks (Ensemble-UNet). Our outcomes on a 3D MRI hippocampus dataset, with and without data augmentation, demonstrate that PNN-UNet outperforms the baseline UNet and several other UNet variants in image segmentation.

**Introduction**

Medical image segmentation using deep learning techniques plays an increasingly crucial role in assisting clinical diagnosis. Every day, hospitals capture exponentially more medical images, making it increasingly difficult to process big data efficiently and effectively. Medical imaging segmentation can be classified into three major categories: 2D, 2.5D, and 3D (Minaee et al., 2021; Zhang et al., 2022). The 2D method is to segment 3D images slice-by-slice, utilizing 2D slices as training and testing data. For the 2.5D category, segmentation algorithms usually segment 3D images slice-by-slice, adding neighboring slices as additional inputs. Lastly, 3D images are cropped and segmented into small cubic images for training and testing. After segmentation, these segmented 2D, 2.5D, or 3D images are then rearranged into their original formats.

It is important to note that different methods have their advantages and disadvantages in 3D medical image segmentation. Nevertheless, researchers have found that the 3D method is not always more accurate than the 2D method in terms of image segmentation of liver, kidney, spleen, pancreas, lung, and brain (Nemoto et al., 2020; Srikrishna et al., 2022; Zettler & Mastmeyer, 2021). From a GPU memory consumption perspective, the 2D method uses a fraction, about one-sixth, of the GPU's memory compared to the 3D method, an attractive attribute in the context of image segmentation (Zettler & Mastmeyer, 2021). Therefore, we chose 2D UNet as the backbone method to demonstrate our ideas in 3D medical image segmentation quickly and effectively.

Even though deep learning holds great promise for medical image processing, widespread adoption remains a work in progress. As LeCun, Bengio, and Hinton (2015) emphasize in their seminal review, the advancement of deep learning depends heavily on the development of novel algorithms and network architectures. Building on our previous work on the planarian neural network (PNN) architecture (Huang et al., 2025), we adapted the PNN in the present study to process three-dimensional (3D) image data. Specifically, we implemented the PNN-UNet to segment the anterior and posterior regions of the hippocampus using a 3D MRI brain dataset. The primary objective of this study is to demonstrate that the PNN-UNet outperforms the original UNet and several of its variants in terms of prediction accuracy for 3D medical image segmentation tasks.

## Methods and Materials

**Datasets**

This study uses a T1-weighted 3D MRI hippocampus dataset to evaluate the PNN-UNet performance. The hippocampus data was obtained from the Medical Segmentation Decathlon (MSD) and acquired from the Vanderbilt University Medical Center, Nashville, US (Antonelli et al., 2022; Simpson et al., 2019). The purpose of this dataset is to segment the anterior and posterior hippocampus from MRI images. The dataset contains 90 healthy adults and 105 adults with a few other brain conditions. A total of 195 individuals are associated with 260 images mapped with 260 masks. The dataset includes 260 3D images and their corresponding masks, both consisting of 9,198 2D slices. The challenge of this dataset is that the regions of interest (ROIs), or the hippocampus, are relatively small objects compared to the size and complexity of the brain tissue.

In this dataset, each 3D image is associated with only one 3D mask or label at a time. Among the hippocampus mask data, there are three types of label values: 0 represents the background, 1 represents the anterior, and 2 represents the posterior (see Table 0.1).

| Mask ID | Mask Name | Pixel-Level Mask Count |
|---|---|---|
| L0 | Background | 15,469,502 |
| L1 | Anterior | 445,938 |
| L2 | Posterior | 410,816 |

Table 0.1. Pixel-Level Hippocampus Mask Count

These 3D images come in different sizes in axial, coronal, and sagittal dimensions, as the corresponding coordinates of x, y, and z. In the source data, the image size in the axial dimension ranges between 31 and 43 pixels. The image size in the coronal dimension ranges between 40 and 59 pixels, and in the sagittal dimension, it ranges between 24 and 47 pixels (see Table 0.2). The differences in dimensions hamper the way to process the segmentation task.

These 3D images need some data preprocessing before being fed into a deep learning model, along with the mask data, according to the modeling strategy associated with this study.

| Measure | X-Axial | Y-Coronal | Z-Sagittal |
|---|---|---|---|
| count | 260 | 260 | 260 |
| mean | 35.376923 | 49.980769 | 35.653846 |
| std | 1.983583 | 3.26021 | 4.218385 |
| min | 31 | 40 | 24 |
| 25% | 34 | 48 | 33 |
| 50% | 35 | 50 | 36 |
| 75% | 37 | 52 | 39 |
| max | 43 | 59 | 47 |

Table 0.2. Size Statistics in X-Axial, Y-Coronal, and Z-Sagittal Dimensions

**Data Preprocessing**

Since our modeling strategy is to use the 2D object segmentation technique in dealing with 3D data, we leave the axial size untouched and extend the size of both the coronal and sagittal dimensions. We did this because the 3D data will be fed into our segmentation model in the 2D (coronal and sagittal dimensions) format along with the axial dimension (see Table 0.3).

| Measure | X-Axial | Y-Coronal | Z-Sagittal |
|---|---|---|---|
| count | 260 | 260 | 260 |
| mean | 35.376923 | 64 | 64 |
| std | 1.983583 | 0 | 0 |
| min | 31 | 64 | 64 |
| 25% | 34 | 64 | 64 |
| 50% | 35 | 64 | 64 |
| 75% | 37 | 64 | 64 |
| max | 43 | 64 | 64 |

Table 0.3. Size Statistics in X-Axial, Y-Coronal, and Z-Sagittal Dimensions After Reshaping

After reshaping, all individual 2D slices became the same size in width and height. Here in our case, we choose 64 as the value of width and height because the baseline model in this study is designed to consume 64x64 images. We randomly selected hippocampus_173 to verify the results of the image and mask size adjustments (see Table 0.4 and Table 0.5).

| Image hippocampus_173 | Before | After |
|---|---|---|
| 12th Slice | 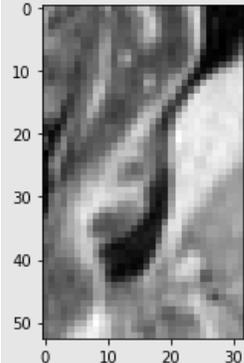 | 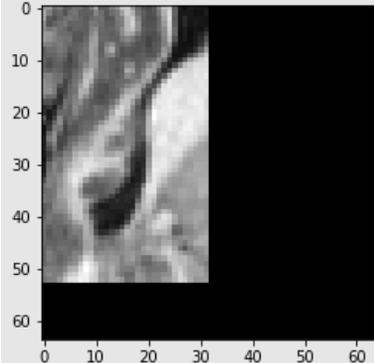 |
| 18th Slice | 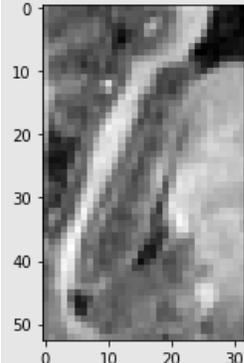 | 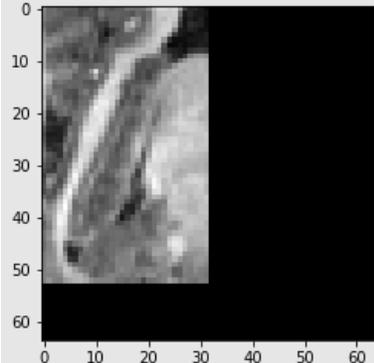 |
| 24th Slice | 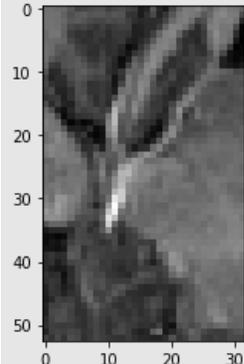 | 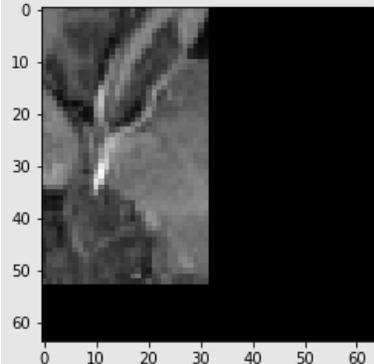 |

Table 0.4. Image Size Adjustment

From the 12th, 18th, and 24th image slices of hippocampus_173, each slice's size was reshaped from 53x32 to 64x64. The black color or value 0 is filled in the expanded areas. The same reshape technique was successfully implemented on the mask slices. In the mask slices, black represents the background, grey represents the anterior hippocampus, and white represents the posterior hippocampus.

| Mask hippocampus_173 | Before | After |
|---|---|---|
| 12th Slice | 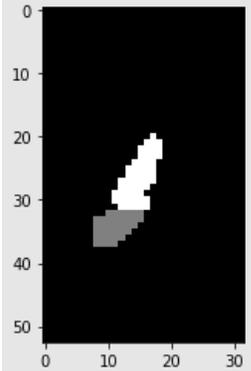 | 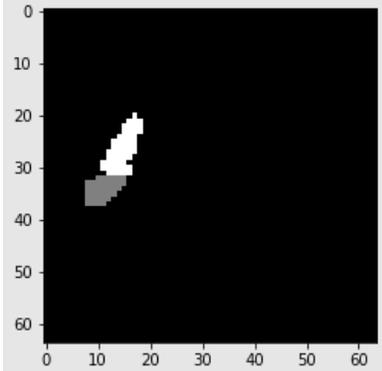 |
| 18th Slice | 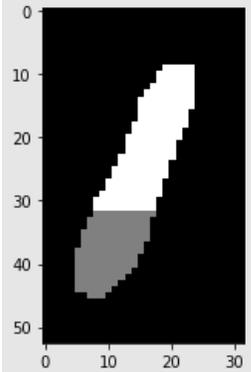 | 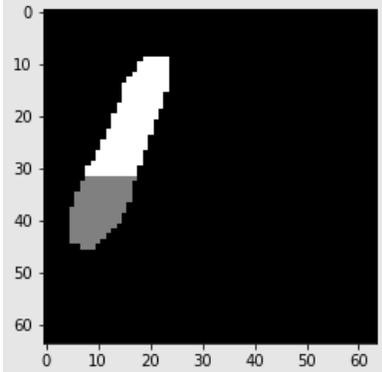 |
| 24th Slice | 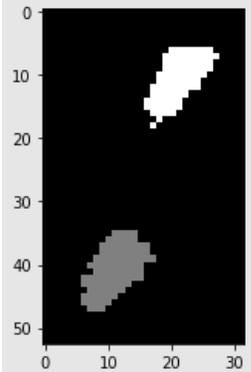 | 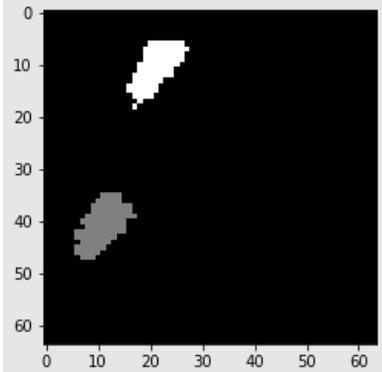 |

Table 0.5. Mask Size Adjustment

**Seeds**

For this study, we designed five random seeds as the five-run method is popular in many deep neural network studies (He et al., 2016a, 2016b; Singh & Cirrone, 2022; Zagoruyko & Komodakis, 2016). Each model is run five times, and each run is associated with a seed. Each seed is assigned an experiment number. This study implemented the five-run method represented by experiment numbers 1, 2, 3, 4, and 5.

It is important to keep in mind that the PyTorch seeds ranges from 0 to 2**64-1 while the Numpy seed range is constrained to 0 and 2**32-1. The smaller range number of Numpy seeding is utilized in this study in order to prevent conflict between PyTorch and Numpy's random number generators. Using the systematic sampling method, our starting seeds were determined by dividing the Numpy seed range by 60 intervals and multiplying the interval by the experiment number. Consequently, the five seeds are 71582788, 143165576, 214748364, 286331153, and 357913941. For a given experiment, the same seed is used for both data loading and model weights initialization.

**Data Split**

As a basis for splitting our dataset, we follow the 60:20:20 principle. A total of 260 3D images are used, of which 60 percent are used for training, 20 percent for validating, and 20 percent for testing (see Figure 0.1).

| | Train Size: 156 | Validation Size: 52 | Test Size: 52 |
|---|---|---|---|
| Seed 1 | Train Dataset | Validation Dataset | Test Dataset |
| Seed 2 | Train Dataset | Validation Dataset | Test Dataset |
| Seed 3 | Train Dataset | Validation Dataset | Test Dataset |
| Seed 4 | Train Dataset | Validation Dataset | Test Dataset |
| Seed 5 | Train Dataset | Validation Dataset | Test Dataset |

Figure 0.1. Data Split Principle

**Data Augmentation**

This study refers to the data augmentation policies in the original UNet paper (Ronneberger et al., 2015). The first data augmentation strategy is elastic deformation (Simard et al., 2003)Secondly, random horizontal flips are applied with p = 0.25 after fine-tuning. These two data augmentation policies are implemented to both images and masks simultaneously in the training dataset. In the validation dataset, only random horizontal flips are applied. The test data remained untouched by design.

**Hardware and Software**

The hippocampus experiments were carried out using Python 3.8.11 and PyTorch 1.12.0 on a workstation running Ubuntu 20.04 and equipped with four RTX 3090 graphics cards.

## Experimental Design

Two phases of this study were conducted: segmentation without data augmentation and segmentation with data augmentation. We examined four neural network architectures in each experimental phase: Deep-UNet, Wide-UNet, Ensemble-UNet, and PNN-UNet. As the first three models are baseline models, PNN-UNet is proposed and tested in this study along with the baseline models. The UNet is the basis of these four models. We chose to use the German Cancer Research Center's implementation of 2D UNet named RecursiveUNet as the backbone architecture because it has flexibility in adjusting contraction and expansion levels (MIC@DKFZ, 2021). Our objective in this study is to compare PNN-UNet to the baseline models in terms of Dice, Jaccard, sensitivity, and specificity. Additionally, all models are run through five random seeds, with the five-run mean serving as the final output.

**Deep-UNet Architecture**

As the default UNet configuration, Deep-UNet has four contractions, four expansions, and an initial filter size of 64 (MIC@DKFZ, 2021). This UNet is called Deep-UNet because it goes down four contractions and up four expansions to segment images (see Figure 0.2), which is deeper than the Wide-UNet configuration discussed next. The number of parameters in Deep-UNet is 31,030,723.

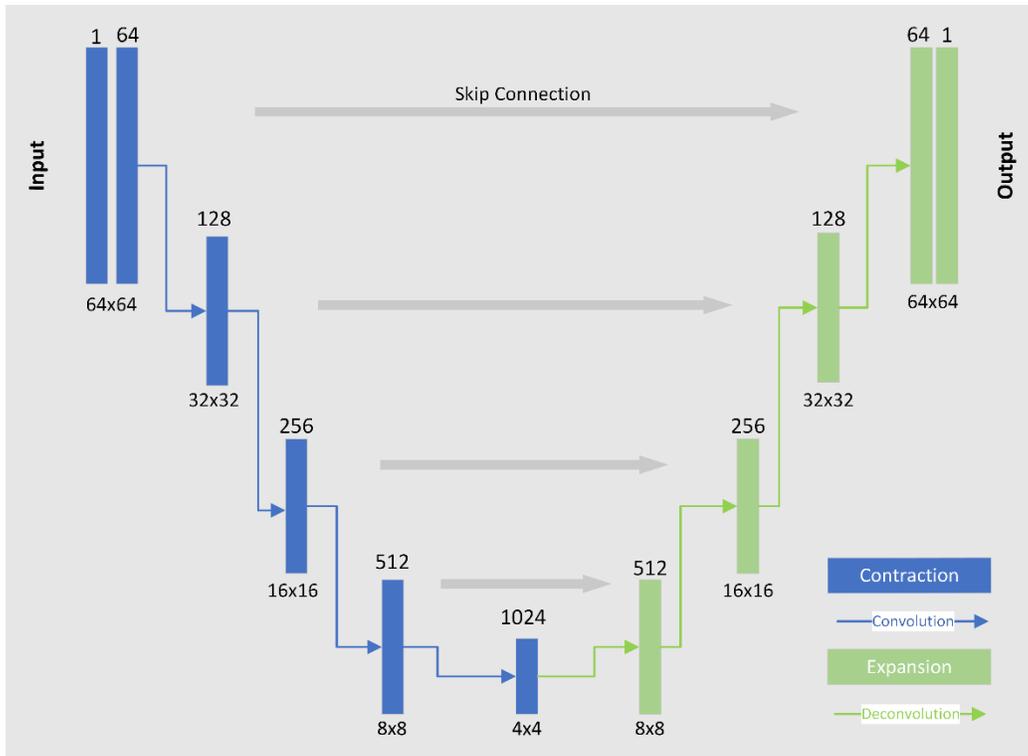

Figure 0.2. Deep-UNet Architecture (MIC@DKFZ, 2021)

**Wide-UNet Architecture**

According to Nguyen et al., (2020), deep networks identify consumer goods better, while wide networks reflect scenes more effectively. In this study, Wide-UNet is configured as a UNet variant. Wide-UNet has only two contractions, two expansions, and an initial filter size of 256 (see Figure 0.3). Compared to Deep-UNet, Wide-UNet goes down by two contractions and then goes up by two expansions with an initial filter size of 256, as opposed to four contractions and expansions with an initial filter size of 64. We set the number of parameters in Wide-UNet to be similar to the number in Deep-UNet, which is 29,762,307. This enabled us to observe that prediction results were more influenced by neural network architecture than the number of parameters. When Deep-UNet and Wide-UNet are combined, we expect enhanced abilities to segment images.

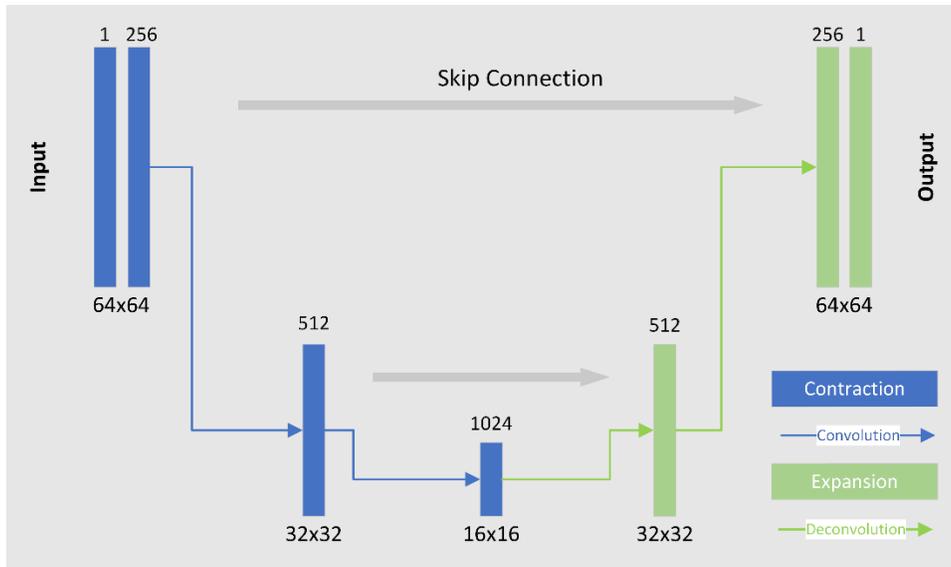

Figure 0.3. Wide-UNet Architecture (MIC@DKFZ, 2021)

**Ensemble-UNet Architecture**

Ensemble-UNet is a stacking algorithm that combines Deep-UNet and Wide-UNet, as depicted in Figure 0.4, using soft voting to merge the results generated by each neural network. There are two proposed training strategies for Ensemble-UNet. The first strategy, called Ensemble-UNet-Transfer (hereafter Ensemble-Transfer), stacks Deep-UNet and Wide-UNet with pre-trained weights to form a transfer learning model connected by soft voting. In this context, "pre-trained" means Deep-UNet and Wide-UNet are trained separately before being stacked together, and their pre-trained weights are loaded under the Ensemble-UNet architecture. Within this transfer learning model, additional training is not required. The second strategy, Ensemble-UNet-Retrain (hereafter Ensemble-Retrain), assembles Deep-UNet and Wide-UNet into an ensemble model without pre-trained weights, connects them with soft voting, and requires retraining upon initializing the Ensemble-UNet. This means Deep-UNet and Wide-UNet are trained together under the Ensemble-UNet architecture. The primary goal of these two

Ensemble-UNet strategies is to determine whether improvements in image segmentation result from transfer learning or retraining.

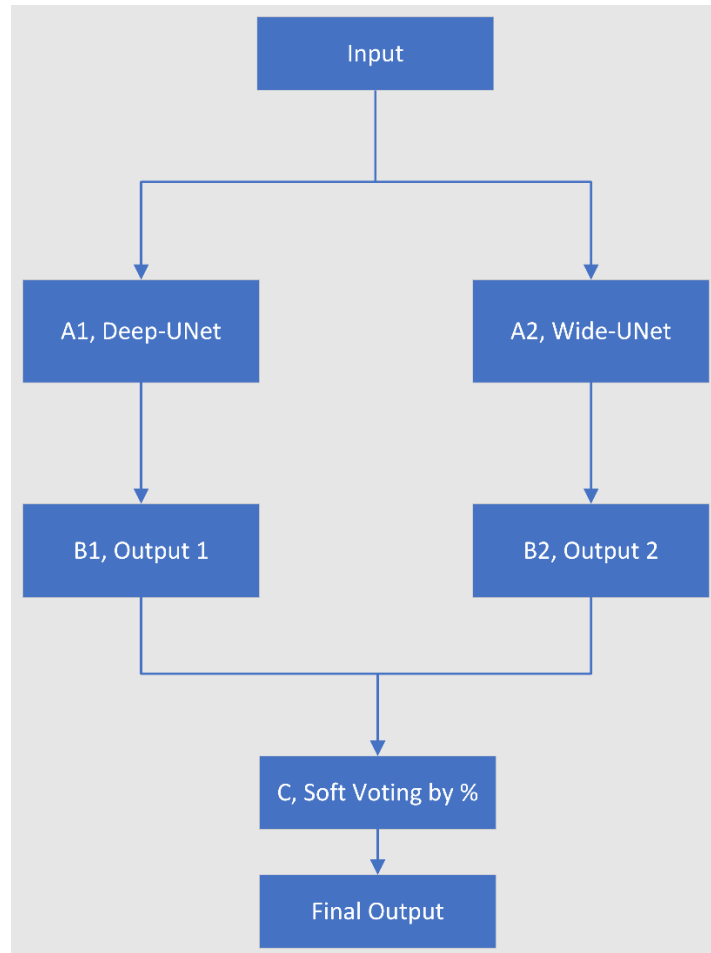

Figure 0.4. Ensemble-UNet Architecture

**PNN-UNet Architecture**

The PNN-UNet is modeled after a planarian's biological nervous system, which is composed of a mass of cephalic ganglions and a pair of ventral nerve (Agata et al., 1998). In the proposed PNN-UNet architecture, a mass of cephalic ganglions is the "autoencoder", the first ventral nerve cord is the "Deep-UNet", and the second ventral nerve cord is the "Wide-UNet". Using the concepts of DenseNet and the classic autoencoder, we created the dense autoencoder (see Figure 0.5). The dense autoencoder aims to capture as many features as possible and coordinate the training process with other components within the PNN framework. The dense autoencoder, Deep-UNet, and Wide-UNet are then ensembled together as the PNN-UNet linked by soft voting (see Figure 0.5 and Figure 0.6).

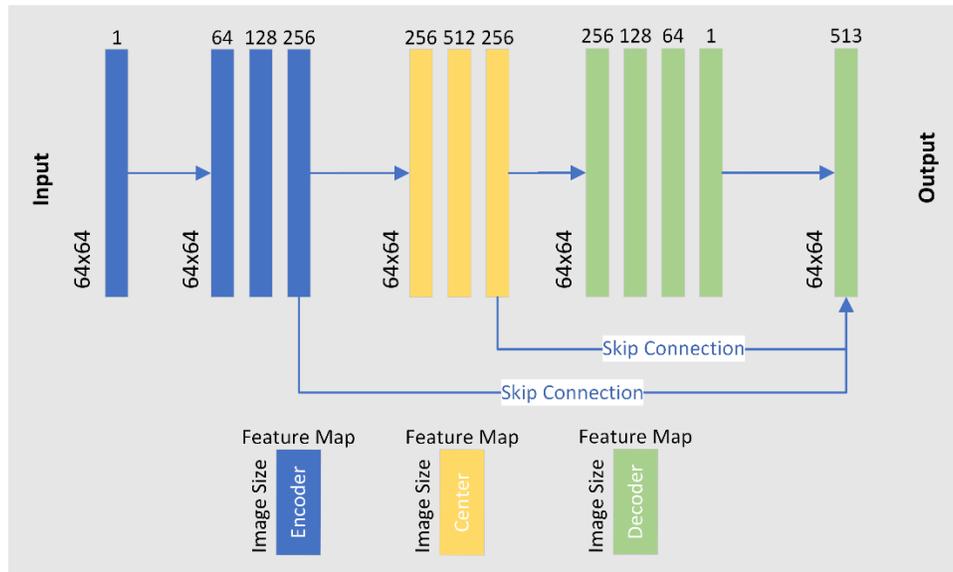

Figure 0.5. Dense Autoencoder Architecture

According to a study of the flatworm nervous system by Reuter and Gustafsson (1995), we proposed adding an artificial cephalic ganglion structure, an autoencoder, before Deep-UNet and Wide-UNet. As Reuter and Gustafsson noted in the research, early bilateria developed a centralized nervous system with a cephalic ganglion. The cephalic ganglion coordinates the frontal sensory receptors on both sides of the body. Our study anticipates that the proposed autoencoder within the PNN framework will facilitate better coordination of input between Deep-UNet and Wide-UNet, as one of its primary functions is feature extraction. Through the forward and backward propagation of PNN-UNet, the input coordination among the autoencoder, Deep-UNet, and Wide-UNet is expected to be enhanced. Consequently, improved input coordination should lead to superior image segmentation.

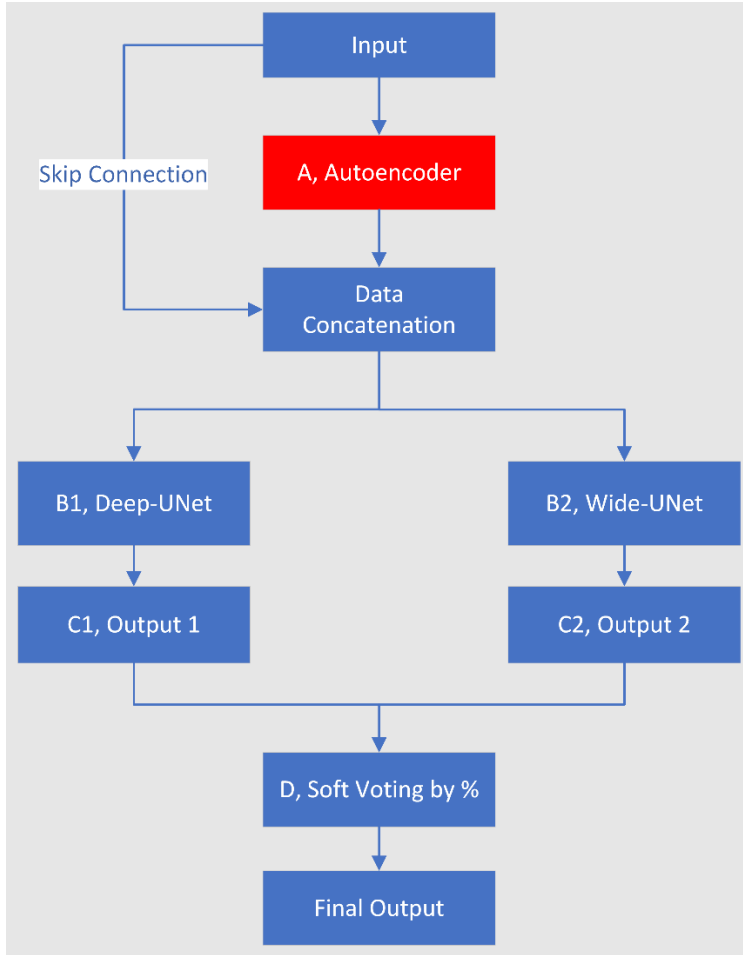

Figure 0.6. PNN-UNet Architecture

**Measurement**

We use four measurements to evaluate the performance of the baseline models and our proposed model, including Dice coefficient, Jaccard index, sensitivity, and specificity.

**Dice.** The Dice score is one of the most popular measures used to evaluate how segmentation models performed (Looman & Campbell, 1960). Dice score is also called Sørensen–Dice index, Sørensen index, F1 score, and Dice's coefficient. The formula for dice score is listed as follows:

$$DSC = \frac{2 \times |X \cap Y|}{|X| + |Y|} = \frac{2TP}{2TP + FP + FN} \qquad (1)$$

In formula (1), X represents predicted segmentation and Y represents the ground truth. The Dice score is to measure how similar X and Y are. The numerator calculates the absolute value of intersection of X and Y and multiplies by 2. Then use the intersection value divided by the absolute value of the sum of X and Y. The range of Dice score is between 0 and 1. The higher the Dice score the more similar the two objects are.

**Jaccard.** The Jaccard index, Jaccard similarity coefficient, or intersection over union (IoU) is another popular statistic in image segmentation to measure the similarity of segmented objects and the ground truth. The formula for Jaccard index is as follows:

$$J(A, B) = \frac{|A \cap B|}{|A \cup B|} = \frac{TP}{TP+FP+FN} \qquad (2)$$

Our study uses formula (2) where A represents prediction and B represents ground truth. The numerator is the absolute value of the intersection of A and B. The denominator is therefore the absolute value of the union of A and B. The Jaccard index is calculated by dividing the intersection of A and B by the union of A and B. This illustrates how closely the predicted segmentation matches the actual segmentation.

**Sensitivity.** The sensitivity metric measures the true positive rate (TPR). A low sensitivity indicates under segmentation of the segmentation model. Having a high sensitivity might be a good sign, but it could also mean that the model does not deal with the negatives well.

$$Sensitivity = \frac{TP}{TP+FN} \qquad (3)$$

The sensitivity is calculated by dividing the true positive (TP) by the sum of the true positive (TP) and false negative (FN). Our segmentation model TP represents correctly predicted hippocampus voxels, while FN represents incorrectly predicted non-hippocampus voxels.

**Specificity.** The specificity metric measures the true negative rate (TNR). A low specificity means segmentation is over segmented. A high specificity could also be a good sign, but it could also indicate the model is not capable of handling positives.

$$Specificity = \frac{TN}{TN+FP} \qquad (4)$$

The specificity is calculated by dividing the true negative (TN) by the sum of the true negative (TN) and false positive (FP). Our segmentation model TN represents correctly predicted non-hippocampus voxels, while FP represents incorrectly predicted hippocampus voxels.

Sensitivity and specificity always work together to ensure model's segmentation performance from both positive and negative aspects.

## Results and Findings

Our findings are presented in two phases. The first phase includes the results without data augmentation, while the second phase encompasses the results with data augmentation. The experimental design comprises four models: Deep-UNet, Wide-UNet, Ensemble-UNet, and PNN-UNet. In Ensemble-UNet, we employed two training strategies, resulting in two models: Ensemble-Transfer and Ensemble-Retrain. Therefore, we have a total of five models for analysis. The results section details the outcomes for all five models. For evaluating these models, we used four measures: Dice, Jaccard, sensitivity, and specificity along with t-tests where $\alpha = 0.05$. The t-test equation is listed as follows:

$$t = \frac{\bar{x} - \mu}{\frac{s}{\sqrt{n}}} \qquad (5)$$

In t-tests, the null hypothesis ($H_0$) holds that the predicted segmentations from the two models have the same mean values based on one of the four measures. Then, the alternative hypothesis ($H_1$) holds that the predicted segmentations from the two models are not equal in terms of mean values based one of on the four measures.

The proposed PNN-UNet model is compared individually to the four baseline models in terms of Dice, Jaccard, sensitivity, and specificity. The predictions involve two segmented objects: the anterior hippocampus (L1) and posterior hippocampus (L2). Each prediction is associated with five seeds. In total, there are five pairs of comparisons across the four measures: PNN-UNet versus Deep-UNet, PNN-UNet versus Wide-UNet, PNN-UNet versus Ensemble-Transfer, PNN-UNet versus Ensemble-Retrain, and Ensemble-Transfer versus Ensemble-Retrain.

**Phase One Results (No Data Augmentation)**

The primary objective of the first phase of this research is to ascertain whether PNN-UNet produces superior segmentation results compared to the baseline models without incorporating augmented data. The outcomes of the five models are illustrated in Figure 0.7.

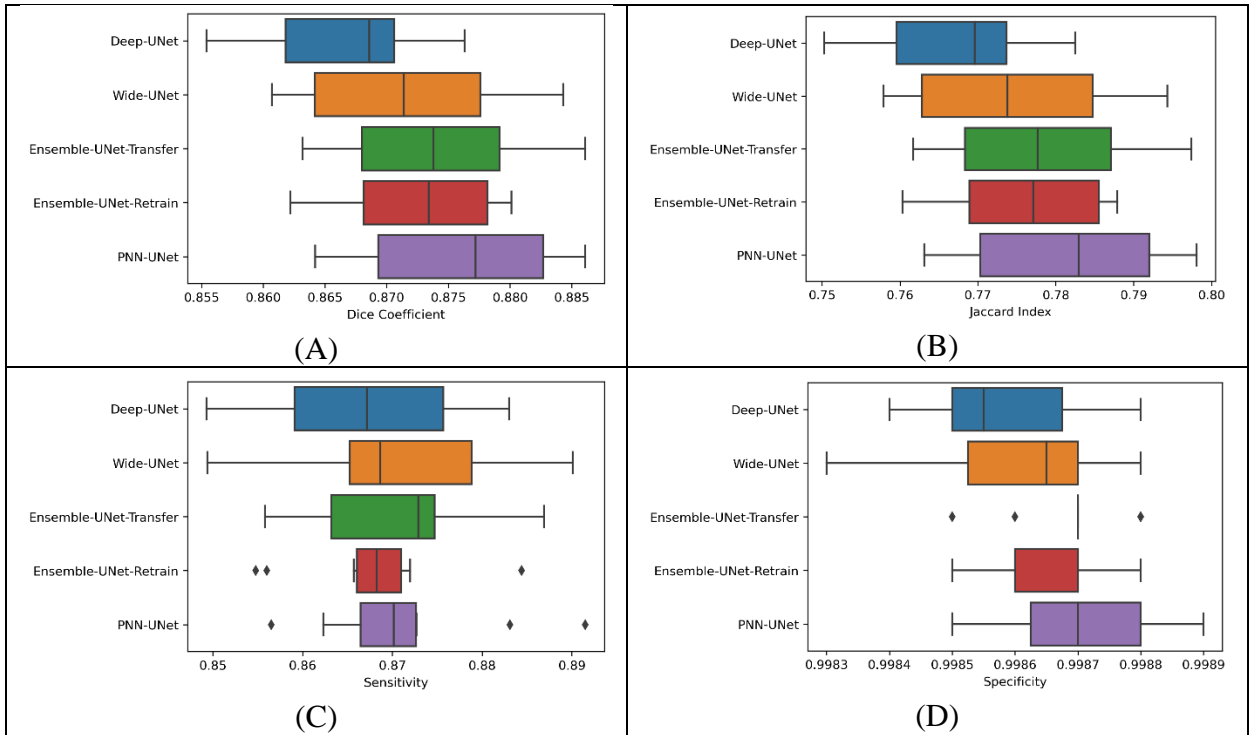

Figure 0.7. Dice, Jaccard, Sensitivity, and Specificity of All Models without Data Augmentation

**Visual Observations.** As observed in Figure 0.7 (A) and (B), the PNN-UNet segment (purple) clearly distinguishes itself from the four other baseline models in terms of Dice and Jaccard scores. Our study evaluated sensitivity and specificity together to assess the model's ability to accurately predict the hippocampus (true positive) and background (true negative). Figure 0.7 (C), does not provide sufficient evidence to determine whether PNN-UNet exhibits higher sensitivity compared to the other models. However, according to Figure 0.7 (D), PNN-UNet demonstrates slightly higher specificity than the other four models.

In an unexpected twist, the Ensemble-UNet model retrained from scratch exhibited inferior performance compared to the Ensemble-UNet with transfer learning. We initially anticipated that the dual-net architecture (Deep-UNet plus Wide-UNet) would yield enhanced segmentation results if we retrained Ensemble-UNet from scratch, as these two networks are jointly trained with a shared loss function and optimizer. However, our expectations were not met.

**Dice.** We conducted a t-test analysis based on each model's prediction results, measured by Dice coefficients without augmented data, as presented in Table 0.6. Three sets of results were generated: L1+L2, L1, and L2. The L1+L2 set represents the t-test outcomes for both anterior and posterior hippocampus segmentation, while the L1 set corresponds to the t-test results for anterior hippocampus segmentation, and the L2 set indicates the t-test results for posterior hippocampus segmentation.

| T-Test in Dice | L1+L2 | L1 | L2 |
|---|---|---|---|
| PNN-UNet versus Deep-UNet | **0.0000 \*\*** | **0.0006 \*\*** | **0.0003 \*\*** |
| PNN-UNet versus Wide-UNet | **0.0001 \*\*** | **0.0187 \*\*** | **0.0066 \*\*** |
| PNN-UNet versus Ensemble-Transfer | **0.0197 \*** | 0.0699 | 0.0765 |
| PNN-UNet versus Ensemble-Retrain | **0.0080 \*\*** | **0.0166 \*** | 0.2114 |
| Ensemble-Transfer versus Ensemble-Retrain | 0.2634 | 0.3706 | 0.6021 |

Table 0.6. T-Test of Dice without Data Augmentation (p-value, α=0.05)

L1+L2 t-test results are generated based on both anterior and posterior hippocampus predictions of the five models measured in Dice. First, PNN-UNet was compared with Deep-UNet, Wide-UNet, Ensemble-Transfer, and Ensemble-Retrain. PNN-UNet's mean Dice coefficient in L1+L2 is 0.8761 compared to Deep-UNet's 0.8665, Wide-UNet's 0.8715, Ensemble-Transfer's 0.8739, and Ensemble-Retrain's 0.8729 (see Table 0.7). The PNN-UNet's L1+L2 Dice is the highest. As shown by the results of the t-test, PNN-UNet differs statistically from Deep-UNet, Wide-UNet, Ensemble-Transfer, and Ensemble-Retrain with the p-values of 0.0000, 0.0001, 0.0197, and 0.0080 (see Table 0.6 - L1+L2).

| Models | L1 Mean | L2 Mean | L1 & L2 Mean |
|---|---|---|---|
| Deep-UNet | 0.871430 | 0.861587 | 0.866509 |
| Wide-UNet | 0.878433 | 0.864615 | 0.871524 |
| Ensemble-Transfer | 0.879716 | 0.868014 | 0.873865 |
| Ensemble-Retrain | 0.878175 | 0.867584 | 0.872880 |
| **PNN-UNet** | **0.883118** | **0.869076** | **0.876097** |

Table 0.7. Five-Run Average of Dice without Data Augmentation

L1 t-test results are generated using the five model predictions of anterior hippocampus measured by Dice. From column L1 in Table 0.6, PNN-UNet is significantly different from Deep-UNet, Wide-UNet, and Ensemble-Retrain, with p-values of 0.0006, 0.0187, and 0.0166 respectively. The associated L1 Dice mean values of PNN-UNet, Deep-UNet, Wide-UNet, and Ensemble-Retrain are 0.8831, 0.8714, 0.8784, and 0.8782, where PNN-UNet in L1 received the highest mean in Dice. PNN-UNet is not statistically significant different from the Ensemble-Transfer model, with p-values of 0.0699 (see Table 0.6 – L1).

L2 t-test results are generated using the five model predictions of posterior hippocampus measured by Dice. From column L2 in Table 0.6, PNN-Net is significantly different from Deep-UNet and Wide-UNet. The p-values of PNN-UNet to Deep-UNet and Wide-UNet are 0.0003 and 0.0066. The associated mean values of L2 prediction for PNN-UNet, Deep-UNet, and Wide-UNet are 0.8691, 0.8616, and 0.8646. However, we cannot define significant differences of L2 predictions of PNN-UNet to the two Ensemble-UNet models. The L2 prediction means of Ensemble-Transfer and Ensemble-Retrain are 0.8680 and 0.8676. Overall, PNN-UNet's L2 prediction mean is the highest among the five models, which is 0.8691 (see Table 0.6 – L2).

Furthermore, we compared Ensemble-Transfer with Ensemble-Retrain using t-test to determine if enhanced Dice may be gained through retraining. This comparison is critical since PNN-UNet uses ensemble learning's stacking logic to combine models. In the case that Ensemble-Transfer and Ensemble-Retrain are statistically identical, that proves that PNN-UNet's improvement on Dice comes from the dense autoencoder and its coordination between Deep-UNet and Wide-UNet. According to the results of the t-test, there is no statistically significant difference between the two Ensemble-UNet models on L1+L2 segmentation. Therefore, it is possible to conclude that the enhancement of PNN-UNet on Dice was due to the addition of the dense autoencoder, as well as the coordination of Deep-UNet and Wide-UNet.

**Jaccard.** We performed t-tests using the prediction results of each model, measured by the Jaccard index without including augmented data. We generated three sets of results for the L1 and L2 masks: L1+L2, L1, and L2. These t-test results are presented in Table 0.8. The L1+L2 set represents t-test outcomes for both anterior and posterior hippocampus segmentation, while the L1 set corresponds to t-test results for the anterior hippocampus segmentation, and the L2 set indicates t-test results for the posterior hippocampus prediction.

| T-Test in Jaccard | L1+L2 | L1 | L2 |
| --- | --- | --- | --- |
| PNN-UNet versus Deep-UNet | **0.0000 \*\*** | **0.0006 \*\*** | **0.0002 \*\*** |
| PNN-UNet versus Wide-UNet | **0.0001 \*\*** | **0.0188 \*** | **0.0051 \*\*** |
| PNN-UNet versus Ensemble-Transfer | **0.0197 \*** | 0.0736 | **0.0433 \*** |
| PNN-UNet versus Ensemble-Retrain | **0.0082 \*\*** | **0.0167 \*** | 0.2114 |
| Ensemble-Transfer versus Ensemble-Retrain | 0.2793 | 0.3751 | 0.6414 |

Table 0.8. T-Test of Jaccard without Data Augmentation (p-value, α=0.05)

L1+L2 t-test results are calculated based on anterior and posterior hippocampus predictions of the five models evaluated in Jaccard (see Table 0.8 – L1+L2 and Table 0.9). As a result of the L1+L2 t-test, PNN-UNet and Deep-UNet have a p-value of 0.0000. Likewise, PNN-UNet and Wide-UNet have a p-value of 0.0001. PNN-UNet and Ensemble-Transfer have a p-value of 0.0197. PNN-UNet and Ensemble-Retrain have a p-value of 0.0082. Thus, PNN-UNet is significantly different from Deep-UNet, Wide-UNet, Ensemble-Transfer, and Ensemble-Retrain with Jaccard index five seeds mean of 0.7815, 0.7668, 0.7744, 0.7781, and 0.7766, where PNN-UNet received the highest mean value of Jaccard.

| Models | L1 Mean | L2 Mean | L1 & L2 Mean |
| --- | --- | --- | --- |
| Deep-UNet | 0.774693 | 0.758984 | 0.766838 |
| Wide-UNet | 0.785448 | 0.763373 | 0.774410 |
| Ensemble-Transfer | 0.787555 | 0.768694 | 0.778125 |
| Ensemble-Retrain | 0.785145 | 0.768114 | 0.776629 |
| **PNN-UNet** | **0.792761** | **0.770329** | **0.781545** |

Table 0.9. Five-Run Average of Jaccard without Data Augmentation

L1 t-test results are generated using the five model predictions of the anterior hippocampus measured by Jaccard. As a result of the L1 t-test, PNN-UNet and Deep-UNet have a p-value of 0.0006. PNN-UNet and Wide-UNet have a p-value of 0.0188. PNN-UNet and Ensemble-Retrain have a p-value of 0.0167. This means that PNN-UNet is significantly different from Deep-UNet, Wide-UNet, and Ensemble-Retrain, with Jaccard index five seed means of 0.7928, 0.7747, 0.7854, and 0.7851, respectively. There is no significant difference in L1 Jaccard between PNN-UNet and Ensemble-Transfer due to the t-test p-value of 0.0736, while Ensemble-Transfer received a five-run average in L1 Jaccard of 0.7876. However, PNN-UNet's Jaccard five-seed mean value of L1 is still the highest among the five models.

The L2 t-test results are generated using the five model predictions for the posterior hippocampus measured by Jaccard. Results from L2 t-tests indicate that PNN-UNet and Deep-UNet have a p-value of 0.0002. PNN-UNet and Wide-UNet have a p-value of 0.0051. PNN-UNet and Ensemble-Transfer have a p-value of 0.0433. Thus, PNN-UNet is statistically different from Deep-UNet, Wide-UNet, and Ensemble-Transfer with Jaccard index seed means of 0.7703, 0.7590, 0.7634, and 0.7687. The t-test p-value of 0.2114 indicates no significant difference between PNN-UNet and Ensemble-Retrain in L2 Jaccard, while Ensemble-Retrain received a five-run average of 0.7681. Among the five models, PNN-UNet still received the highest L2 Jaccard index score.

Additionally, we compared Ensemble-Transfer to Ensemble-Retrain followed by a t-test to determine if retraining could improve Jaccard. In Table 0.8's Ensemble-Transfer versus Ensemble-Retrain, the t-test values of L1+L2, L1, and L2 are all larger than 0.05. The mean values of L1+L2, L1, and L2 segmentation generated by Ensemble-Transfer are 0.7781, 0.7876, and 0.7687. Therefore, the mean values of L1+L2, L1, and L2 segmentation generated by Ensemble-Ensemble are 0.7781, 0.7876, and 0.7687. That indicates Ensemble-Transfer and Ensemble-Retrain do not differ from each other.

**Sensitivity.** By measuring the sensitivity of each model's prediction results, we calculated t-tests. In Table 0.10, the results of t-tests are presented in three sets: L1+L2, L1, and L2. Measured in sensitivity, L1+L2 represents anterior and posterior hippocampus segmentation t-test results, L1 represents anterior hippocampus prediction t-test results, and L2 represents posterior hippocampus prediction t-test results.

| T-Test in Sensitivity | L1+L2 | L1 | L2 |
|---|---|---|---|
| PNN-UNet versus Deep-UNet | 0.2535 | 0.6735 | 0.2010 |
| PNN-UNet versus Wide-UNet | 0.8560 | 0.9081 | 0.9056 |
| PNN-UNet versus Ensemble-Transfer | 0.6833 | 0.8257 | 0.7295 |
| PNN-UNet versus Ensemble-Retrain | 0.1573 | 0.1770 | 0.6383 |
| Ensemble-Transfer versus Ensemble-Retrain | 0.2255 | 0.1671 | 0.8122 |

Table 0.10. T-Test of Sensitivity without Data Augmentation (p-value, $\alpha=0.05$)

According to Table 0.10, PNN-UNet's t-test results do not show significant differences to Deep-UNet, Wide-UNet, Ensemble-Transfer, and Ensemble-Retrain in terms of L1+L2, L1, and L2. There is also no statistically significant difference between the Ensemble-Transfer and Ensemble-Retrain in L1+L2, L1, and L2.

| Models | L1 Mean | L2 Mean | L1 & L2 Mean |
| --- | --- | --- | --- |
| Deep-UNet | 0.874259 | 0.859992 | 0.867125 |
| Wide-UNet | 0.876419 | 0.864821 | 0.870620 |
| Ensemble-Transfer | 0.876152 | 0.864533 | 0.870342 |
| Ensemble-Retrain | 0.871828 | 0.863770 | 0.867799 |
| PNN-UNet | **0.877029** | **0.865413** | **0.871221** |

Table 0.11. Five-Run Average of Sensitivity without Data Augmentation

In terms of segmentation sensitivity, PNN-UNet is slightly better than Deep-UNet, Wide-UNet, Ensemble-Transfer, and Ensemble-Retrain with the highest L1 five-run average of 0.877029, the highest L2 five-run average of 0.865413, and the highest L1 and L2 five-run average of 0.871221 (see Table 0.11). This minor improvement in sensitivity demonstrates PNN-UNet's improved ability to work with the true positives.

**Specificity.** Using t-tests, we determined the specificity of each model's prediction results. Table 0.12 presents three sets of t-test results: L1+L2, L1, and L2. Specificity is measured by L1+L2 for anterior and posterior hippocampus segmentation, L1 for anterior hippocampus prediction, and L2 for posterior hippocampus prediction.

| T-Test in Specificity | L1+L2 | L1 | L2 |
| --- | --- | --- | --- |
| PNN-UNet versus Deep-UNet | **0.0090 \*\*** | **0.0273 \*** | 0.1918 |
| PNN-UNet versus Wide-UNet | **0.0419 \*** | 0.0899 | 0.2798 |
| PNN-UNet versus Ensemble-Transfer | 0.1033 | **0.0263 \*** | 0.8342 |
| PNN-UNet versus Ensemble-Retrain | 0.2237 | 0.1267 | 0.8641 |
| Ensemble-Transfer versus Ensemble-Retrain | 0.3924 | 0.0609 | 0.9546 |

Table 0.12. T-Test of Specificity without Data Augmentation (p-value, α=0.05)

The L1+L2 t-test results indicate that PNN-UNet differs significantly from Deep-UNet and Wide-UNet with p-values of 0.0090 and 0.0419. Additionally, PNN-UNet, Ensemble-Transfer, and Ensemble-Retrain do not show significant differences. The L1 t-test results suggest that PNN-UNet differs significantly from Deep-UNet and Ensemble-Transfer with p-values of 0.0273 and 0.0263. According to the L2 t-test results, there are no significant differences between the five models with p-values of 0.1918, 0.2798, 0.8342, 0.8641, and 0.9546.

| Models | L1 Mean | L2 Mean | L1 & L2 Mean |
|---|---|---|---|
| Deep-UNet | 0.998562 | 0.998589 | 0.998576 |
| Wide-UNet | 0.998646 | 0.998558 | 0.998602 |
| Ensemble-Transfer | 0.998684 | 0.998653 | 0.998669 |
| Ensemble-Retrain | 0.998707 | 0.998655 | 0.998681 |
| PNN-UNet | **0.998761** | **0.998661** | **0.998711** |

Table 0.13. Five-Run Average of Specificity without Data Augmentation

From Table 0.13, we could observe that PNN-UNet's five-run average of L1, L2, and L1 & L2 are the highest among all the models with slight improvements over the other models. That means PNN-UNet has a slightly better ability to deal with the true negatives.

**Phase Two Results (With Data Augmentation)**

Phase two is almost the same as phase one, except that during training and validation, data augmentation strategies are used. Five models are included: Deep-UNet, Wide-UNet, Ensemble-Transfer, Ensemble-Retrain, and PNN-UNet. Dice, Jaccard, sensitivity, and specificity are evaluated along with t-tests for significant analysis. Overall, we have observed a few different patterns when data augmentation is involved compared to when there is no data augmentation.

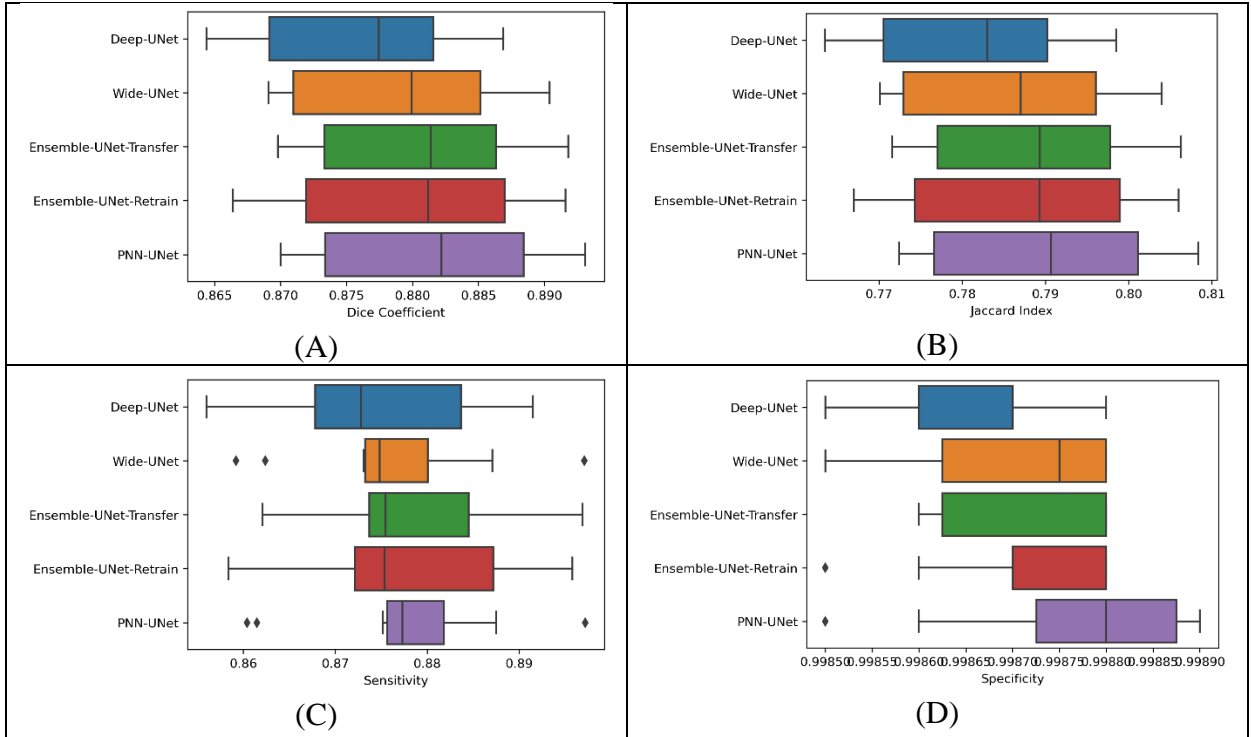

Figure 0.8. Dice, Jaccard, Sensitivity, and Specificity of All Models with Data Augmentation

**Visual Observations.** According to Figure 0.8 (A), (B), and (D), the PNN-UNet (purple) shows some superiority over the other baseline models in Dice, Jaccard, and specificity. Figure 0.8 (D) depicts no difference between PNN-UNet (purple) and other models in terms of sensitivity.

Furthermore, Ensemble-Transfer and Ensemble-Retrain do not differ from our visual observations of Figure 0.8. Therefore, Ensemble-Retrain does not gain additional benefits in Dice, Jaccard, sensitivity, and specificity. In the following sections, we use Dice, Jaccard, sensitivity, and specificity along with t-tests to verify the findings from the visual observations.

**Dice.** We conducted a t-test analysis on the prediction results of each model, measured by Dice coefficients with augmented data (see Table 0.14). The analysis produced three sets of results: L1+L2, L1, and L2. Table 0.14 presents the t-test outcomes for the segmentation of the anterior and posterior hippocampus. L1+L2 represents the t-test outcomes derived from the combined L1 and L2 masks, which encompass both the anterior and posterior hippocampus regions. The L1 results refer to the t-test outcomes specific to the anterior hippocampus (L1) segmentation, while the L2 results pertain to the t-test outcomes for the posterior hippocampus (L2) segmentation.

| T-Test in Dice | L1+L2 | L1 | L2 |
|---|---|---|---|
| PNN-UNet versus Deep-UNet | **0.0000 \*\*** | **0.0001 \*\*** | **0.0110 \*** |
| PNN-UNet versus Wide-UNet | **0.0000 \*\*** | **0.0000 \*\*** | **0.0237 \*** |
| PNN-UNet versus Ensemble-Transfer | **0.0241 \*** | **0.0034 \*\*** | 0.5394 |
| PNN-UNet versus Ensemble-Retrain | **0.0019 \*\*** | **0.0236 \*** | **0.0498 \*** |
| Ensemble-Transfer versus Ensemble-Retrain | 0.3498 | 0.3157 | 0.1730 |

Table 0.14. T-Test of Dice with Data Augmentation (p-value, α=0.05)

In Table 0.14, the L1+L2 column reveals that the t-test results demonstrate a significant difference between PNN-UNet and all baseline models. Similarly, in the L1 column, PNN-UNet exhibits significant differences from the baseline models. However, in the L2 column, PNN-UNet does not display a significant difference when compared to the Ensemble-Transfer model. For the remaining baseline models in the L2 column, significant differences are observed when compared to PNN-UNet.

In Table 0.14 another notable comparison is between Ensemble-Transfer and Ensemble-Retrain. Examining the columns for L1+L2, L1, and L2 reveals that there is no significant difference between Ensemble-Transfer and Ensemble-Retrain. This suggests that retraining the ensembled Deep-UNet and Wide-UNet does not lead to an increase in segmentation accuracy under the influence of augmented data.

| Models | L1 Mean | L2 Mean | L1 & L2 Mean |
|---|---|---|---|
| Deep-UNet | 0.882820 | 0.869494 | 0.876157 |
| Wide-UNet | 0.885915 | 0.872098 | 0.879007 |
| Ensemble-Transfer | 0.887330 | 0.873757 | 0.880543 |
| Ensemble-Retrain | 0.887669 | 0.872492 | 0.880081 |
| PNN-UNet | **0.888897** | **0.874161** | **0.881529** |

Table 0.15. Five-Run Average of Dice with Data Augmentation

Table 0.15 listed the average Dice scores of the predicted anterior hippocampus (L1), posterior hippocampus, and the L1+ L2 segmentations. PNN-UNet's L1, L2, and L1+L2 are the highest compared to the baseline models in terms of Dice.

**Jaccard.** We performed t-tests using each model's prediction results measured by the Jaccard index with augmented data, generating three sets of results for L1 and L2 masks: L1+L2, L1, and L2. Table 0.16 displays the t-test results, with L1+L2 representing the combined results for anterior and posterior hippocampus segmentation. L1 corresponds to the anterior hippocampus segmentation, while L2 refers to the posterior hippocampus segmentation.

| T-Test in Jaccard | L1+L2 | L1 | L2 |
|---|---|---|---|
| PNN-UNet versus Deep-UNet | **0.0000 \*\*** | **0.0001 \*\*** | **0.0102 \*** |
| PNN-UNet versus Wide-UNet | **0.0000 \*\*** | **0.0000 \*\*** | **0.0151 \*** |
| PNN-UNet versus Ensemble-Transfer | **0.0252 \*** | **0.0037 \*\*** | 0.5685 |
| PNN-UNet versus Ensemble-Retrain | **0.0024 \*\*** | **0.0281 \*** | 0.0582 |
| Ensemble-Transfer versus Ensemble-Retrain | 0.4024 | 0.2506 | 0.1729 |

Table 0.16. T-Test of Jaccard with Data Augmentation (p-value, α=0.05)

In Table 0.16, the columns L1+L2 and L1 show that PNN-UNet has significant differences compared to the baseline models. In column L2, PNN-UNet demonstrates significant differences with respect to Deep-UNet and Wide-UNet, but not with the two ensemble models. Additionally, there is no significant difference between Ensemble-Transfer and Ensemble-Retrain in terms of the Jaccard index when considering the influence of augmented data.

| Models | L1 Mean | L2 Mean | L1 & L2 Mean |
|---|---|---|---|
| Deep-UNet | 0.792086 | 0.770860 | 0.781473 |
| Wide-UNet | 0.797074 | 0.774826 | 0.785950 |
| Ensemble-Transfer | 0.799360 | 0.777498 | 0.788429 |
| Ensemble-Retrain | 0.799970 | 0.775667 | 0.787819 |
| PNN-UNet | **0.801815** | **0.778057** | **0.789936** |

Table 0.17. Five-Run Average of Jaccard with Data Augmentation

Table 0.17 presents the average Jaccard index values for the L1, L2, and combined L1+L2 segmentations across all five models. PNN-UNet outperforms the other models with a five-run average Jaccard index of 0.801815 for L1, 0.778057 for L2, and 0.789936 for the combined L1+L2 segmentations.

**Sensitivity.** We calculated t-tests by measuring each model's sensitivity to prediction results with the influence from the augmented data (see Table 0.18). The t-test results are presented in three sets: L1+L2, L1, and L2. In terms of sensitivity, L1+L2 represents the t-test results for the combined anterior and posterior hippocampus segmentations, L1 corresponds to the t-test results for the anterior hippocampus predictions, and L2 represents the t-test results for the posterior hippocampus predictions.

| T-Test in Sensitivity | L1+L2 | L1 | L2 |
|---|---|---|---|
| PNN-UNet versus Deep-UNet | 0.0636 | 0.1912 | 0.2520 |
| PNN-UNet versus Wide-UNet | **0.0284 *** | 0.0910 | 0.2452 |
| PNN-UNet versus Ensemble-Transfer | 0.7187 | 0.9473 | 0.6645 |
| PNN-UNet versus Ensemble-Retrain | 0.8297 | 0.5020 | **0.0103 *** |
| Ensemble-Transfer versus Ensemble-Retrain | 0.7148 | 0.5232 | **0.0325 *** |

Table 0.18. T-Test of Sensitivity with Data Augmentation (p-value, α=0.05)

As per Table 0.18, the t-test results highlight a significant difference between PNN-UNet and Wide-UNet in L1+L2, as well as between PNN-UNet and Ensemble-Retrain in L2. Apart from these differences, there are no other significant disparities among the models compared with PNN-UNet. Interestingly, a noticeable difference exists between Ensemble-Transfer and Ensemble-Retrain in L2 segmentation, with a p-value less than 0.05, signifying their significant distinction. Conversely, in L1+L2 and L1, there are no observable differences between Ensemble-Transfer and Ensemble-Retrain when considering the impact of augmented data on sensitivity.

| Models | L1 Mean | L2 Mean | L1 & L2 Mean |
|---|---|---|---|
| Deep-UNet | 0.881924 | 0.867634 | 0.874779 |
| Wide-UNet | 0.882515 | 0.869345 | 0.875930 |
| Ensemble-Transfer | 0.884491 | **0.871075** | **0.877783** |
| Ensemble-Retrain | **0.886263** | 0.868181 | 0.877222 |
| PNN-UNet | 0.884414 | 0.870653 | 0.877534 |

Table 0.19. Five-Run Average of Sensitivity with Data Augmentation

Table 0.19 presents the five-run average values for the five models concerning anterior hippocampus (L1) prediction, posterior hippocampus (L2) prediction, and combined anterior and posterior (L1+L2) prediction with augmented data. Ensemble-Retrain achieves the highest L1 five-run average, while Ensemble-Transfer attains the highest five-run averages for both L2 and L1+L2 predictions. In terms of sensitivity, PNN-UNet falls between Ensemble-Transfer and Ensemble-Retrain.

**Specificity.** Under the impact of augmented data, we employed t-tests to evaluate the specificity of each model's predictions. In Table 0.20, three sets of t-test results are presented: L1+L2, L1, and L2. The specificity of the model is measured by L1+L2 for anterior and posterior hippocampus segmentation, by L1 for anterior hippocampus prediction, and by L2 for posterior hippocampus prediction.

| T-Test in Specificity | L1+L2 | L1 | L2 |
|---|---|---|---|
| PNN-UNet versus Deep-UNet | **0.0004 \*\*** | **0.0088 \*\*** | **0.0221 \*** |
| PNN-UNet versus Wide-UNet | **0.0184 \*** | **0.0094 \*\*** | 0.3319 |
| PNN-UNet versus Ensemble-Transfer | **0.0177 \*** | **0.0299 \*** | 0.3287 |
| PNN-UNet versus Ensemble-Retrain | 0.1957 | 0.1832 | 0.9359 |
| Ensemble-Transfer versus Ensemble-Retrain | 0.9013 | 0.6685 | 0.7142 |

Table 0.20. T-Test of Specificity with Data Augmentation (p-value, α=0.05)

Table 0.20 indicates that PNN-UNet exhibits a significant difference in terms of specificity when compared to Deep-UNet, Wide-UNet, and Ensemble-Transfer for both L1+L2 and L1. However, in L2, PNN-UNet only shows a significant difference when compared to Deep-UNet. The remaining t-test results do not show any significant differences. Additionally, the results show no statistical differences in terms of L1+L2, L1, and L2 predictions between Ensemble-Transfer and Ensemble-Retrain, implying that retraining ensemble models does not result in improved specificity with augmented data.

| Models | L1 Mean | L2 Mean | L1 & L2 Mean |
|---|---|---|---|
| Deep-UNet | 0.998675 | 0.998647 | 0.998661 |
| Wide-UNet | 0.998740 | 0.998684 | 0.998712 |
| Ensemble-Transfer | 0.998752 | 0.998702 | 0.998727 |
| Ensemble-Retrain | 0.998737 | 0.998712 | 0.998724 |
| PNN-UNet | **0.998792** | **0.998714** | **0.998753** |

Table 0.21. Five-Run Average of Specificity with Data Augmentation

In Table 0.21, it can be observed that PNN-UNet achieved the highest specificity in L1, L2, and L1+L2, surpassing all the other models. Specifically, PNN-UNet's specificity values were 0.998792 in L1, 0.998714 in L2, and 0.998753 in L1+L2, indicating enhanced background regions identification in images.

## Discussion

The PNN-UNet demonstrated improvement overall in all the measures, with Dice and Jaccard having statistically significant improvement. In order to figure out how Dice and Jaccard improved so significantly while sensitivity and specificity did not, we decided to analyze the results generated by the four measurements.

The improved sensitivity means increased TP and reduced FN. We mark an upward arrow sign next to TP and a downward arrow sign next to FN as displayed in formula (6).

$$Sensitivity = \frac{TP\uparrow}{TP\uparrow + FN\downarrow} \quad (6)$$

Our experiments on sensitivity involved two scenarios: the first phase without data augmentation and the second phase with data augmentation. In phase one of our experiments, we did not observe significant improvements in sensitivity. However, it is noteworthy that PNN-UNet had the highest sensitivity results among the five models (Table 0.11 and Table 0.12). While we observed a moderate improvement in the sensitivity of PNN-UNet, it was not statistically significant. In phase two, the sensitivity results ranked approximately second among the five models, but L1+L2 sensitivity is superior to the Wide-UNet model and L2 sensitivity is superior to the Ensemble-Retrain model. The sensitivity of PNN-UNet in phase two is no different from that of the ranked first model (Table 0.18 and Table 0.19). PNN-UNet did not earn significant differences in phase one but earned two significant differences in phase two.

As with sensitivity, improved specificity results in increased TN and reduced FP. We mark an upward arrow next to TN and a downward arrow sign next to FP as displayed in formula (7).

$$Specificity = \frac{TN\uparrow}{TN\uparrow + FP\downarrow} \quad (7)$$

The specificity is also evaluated with two different scenarios: models trained without data augmentation in phase one and models trained with data augmentation in phase two. In phase one, PNN-UNet's specificity is the highest among the five models. The significant differences of PNN-UNet happened only to Deep-UNet in L1+L2 and L1, Wide-UNet in L1+L2, and Ensemble-Transfer in L1 (Table 0.12 and Table 0.13). The specificity of PNN-UNet ranked first among the five models in phase two as well. Therefore, PNN-UNet differed significantly from Deep-UNet in L1+L2, L1, and L2, Wide-UNet in L1+L2 and L1, and Ensemble-Transfer in L1+L2 and L1 (Table 0.20 and Table 0.21). Overall, there were four significant differences in phase one of PNN-UNet, and seven in phase two.

The upward and downward arrows from formulas (6) and (7) lead us to review the formulas for Dice coefficient and Jaccard index. We add a few upward arrows and downward arrows to the Dice coefficient formula (8) and Jaccard index formula (9).

$$DSC = \frac{2TP\uparrow}{2TP\uparrow + FP\downarrow + FN\downarrow} \quad (8)$$

$$J(A,B) = \frac{TP\uparrow}{TP\uparrow + FP\downarrow + FN\downarrow} \quad (9)$$

The upward and downward arrows in formulas (8 and 9) allowed us to identify the cause of the significant improvement in Dice and Jaccard. Despite the slight improvements in sensitivity and specificity, PNN-UNet significantly improved Dice and Jaccard either with or without data augmentation according to our significant test results. Using the Dice and Jaccard formulas with the upward and downward arrows, we see both an increase in the numerator and a decrease in the denominator. The double impact of the numerator and denominator accelerates PNN-UNet's Dice and Jaccard progress toward significance.

In terms of artificial intelligence (AI), we were able to generate better segmentation using the PNN-UNet with the dense autoencoder as its brain, Deep-UNet and Wide-UNet as its dual nerve cords by mimicking the biological nervous system structure of planarians. The performance of segmentation is not improved by stacking two neural networks together and retraining the ensembled model. Our Ensemble-Transfer and Ensemble-Retrain experiments indicate that the segmentation performance between these two methods do not differ significantly with each other. The Ensemble-UNet comparisons on the contrary indicate that the newly added autoencoder with skip connections in PNN-UNet have made the difference.

Nguyen et al.'s paper (2020) regarding deep and wide networks stated that deep architecture is better at identifying consumer goods (objects) and wide architecture is better at identifying scenes (background), such as seashores, libraries, and bookshops. We have incorporated both deep and wide architecture into PNN-UNet, which consists of a Deep-UNet, a Wide-UNet, and an additional autoencoder that emulates planarian's brain to coordinate the two neural networks. Our experimental results show that both object and background predictions have improved in segmenting hippocampus. In other words, our architecture can leverage the benefits of both deep and wide networks, in addition to the autoencoder.

There are a few potential developments for PNN architecture in the future. From a deep learning perspective, the study of PNN architecture could be extended into fields such as semi-supervised learning, unsupervised learning, and reinforcement learning. From a data source perspective, PNN architecture could be applied to different kinds of datasets, such as liver, heart, pancreas, hepatic vessels, and many others. In terms of architecture design, the PNN architecture can be extended to use different kinds of biological nerve systems' structures to meet a variety of prediction requirements.

## Conclusion

In conclusion, when segmenting the 3D MRI hippocampus dataset, PNN-UNet outperforms all baseline models mentioned in this study, including Deep-UNet, Wide-UNet, Ensemble-Transfer, and Ensemble-Retrain, either with or without data augmentation. A variety of measures were used to measure the performance of PNN-UNet and the baseline model, including Dice, Jaccard, sensitivity, and specificity. Dice and Jaccard of PNN-UNet both demonstrated notable improvements in segmenting the anterior and posterior hippocampus. In terms of sensitivity and specificity, PNN-UNet showed moderate improvements. The results from our experiments revealed that PNN-UNet achieved significant improvements in both Dice and Jaccard scores, indicating a higher level of accuracy in segmenting objects from backgrounds. Furthermore, PNN-UNet demonstrated moderate improvements in sensitivity and specificity, which further supports its ability to distinguish between foreground and background regions. Overall, these results suggest that PNN-UNet is a promising framework for image segmentation tasks, as it offers substantial improvements in performance over the baseline models.

Reference

Agata, K., Soejima, Y., Kato, K., Kobayashi, C., Umesono, Y., & Watanabe, K. (1998). Structure of the planarian central nervous system (CNS) revealed by neuronal cell markers. *Zoological Science*, *15*(3), 433–440.

Antonelli, M., Reinke, A., Bakas, S., Farahani, K., Kopp-Schneider, A., Landman, B. A., Litjens, G., Menze, B., Ronneberger, O., Summers, R. M., & others. (2022). The medical segmentation decathlon. *Nature Communications*, *13*(1), 4128.

He, K., Zhang, X., Ren, S., & Sun, J. (2016a). Deep residual learning for image recognition. *Proceedings of the IEEE Conference on Computer Vision and Pattern Recognition*, 770–778.

He, K., Zhang, X., Ren, S., & Sun, J. (2016b). Identity mappings in deep residual networks. *Computer Vision–ECCV 2016: 14th European Conference, Amsterdam, The Netherlands, October 11–14, 2016, Proceedings, Part IV 14*, 630–645.

Huang, Z., Newman, M., Vaida, M., Bellur, S., Sadeghian, R., Siu, A., ... & Huggins, K. (2025). Planarian Neural Networks: Evolutionary Patterns from Basic Bilateria Shaping Modern Artificial Neural Network Architectures. arXiv preprint arXiv:2501.04700.

LeCun, Y., Bengio, Y., & Hinton, G. (2015). Deep learning. *Nature*, *521*(7553), 436–444.

Looman, J., & Campbell, J. (1960). Adaptation of Sorensen's K (1948) for estimating unit affinities in prairie vegetation. *Ecology*, *41*(3), 409–416.

MIC@DKFZ. (2021). *Basic UNet Example*. GitHub. https://github.com/MIC-DKFZ/basic_unet_example

Minaee, S., Boykov, Y. Y., Porikli, F., Plaza, A. J., Kehtarnavaz, N., & Terzopoulos, D. (2021). Image segmentation using deep learning: A survey. *IEEE Transactions on Pattern Analysis and Machine Intelligence*.

Nemoto, T., Futakami, N., Yagi, M., Kumabe, A., Takeda, A., Kunieda, E., & Shigematsu, N. (2020). Efficacy evaluation of 2D, 3D U-Net semantic segmentation and atlas-based segmentation of normal lungs excluding the trachea and main bronchi. *Journal of Radiation Research*, *61*(2), 257–264.

Nguyen, T., Raghu, M., & Kornblith, S. (2020). Do wide and deep networks learn the same things? Uncovering how neural network representations vary with width and depth. *ArXiv Preprint ArXiv:2010.15327*.

Reuter, M., & Gustafsson, M. (1995). The flatworm nervous system: Pattern and phylogeny. *The Nervous Systems of Invertebrates: An Evolutionary and Comparative Approach: With a Coda Written by TH Bullock*, 25–59.

Ronneberger, O., Fischer, P., & Brox, T. (2015). U-net: Convolutional networks for biomedical image segmentation. *Medical Image Computing and Computer-Assisted Intervention–MICCAI 2015: 18th International Conference, Munich, Germany, October 5-9, 2015, Proceedings, Part III 18*, 234–241.

Simard, P. Y., Steinkraus, D., Platt, J. C., & others. (2003). Best practices for convolutional neural networks applied to visual document analysis. *Icdar*, *3*(2003).

Simpson, A. L., Antonelli, M., Bakas, S., Bilello, M., Farahani, K., Van Ginneken, B., Kopp-Schneider, A., Landman, B. A., Litjens, G., Menze, B., & others. (2019). A large annotated medical image dataset for the development and evaluation of segmentation algorithms. *ArXiv Preprint ArXiv:1902.09063*.

Singh, P., & Cirrone, J. (2022). A Data-Efficient Deep Learning Framework for Segmentation and Classification of Histopathology Images. *ArXiv Preprint ArXiv:2207.06489*.

Srikrishna, M., Heckemann, R. A., Pereira, J. B., Volpe, G., Zettergren, A., Kern, S., Westman, E., Skoog, I., & Schöll, M. (2022). Comparison of two-dimensional-and three-dimensional-based U-Net architectures for brain tissue classification in one-dimensional brain CT. *Frontiers in Computational Neuroscience*, *15*, 123.

Zagoruyko, S., & Komodakis, N. (2016). Wide Residual Networks. *CoRR*, *abs/1605.07146*. http://arxiv.org/abs/1605.07146

Zettler, N., & Mastmeyer, A. (2021). Comparison of 2D vs. 3D U-Net Organ Segmentation in abdominal 3D CT images. *ArXiv Preprint ArXiv:2107.04062*.

Zhang, Y., Liao, Q., Ding, L., & Zhang, J. (2022). Bridging 2D and 3D segmentation networks for computation-efficient volumetric medical image segmentation: An empirical study of 2.5 D solutions. *Computerized Medical Imaging and Graphics*, *99*, 102088.